\documentclass[11pt]{report}

\makeatletter
\renewcommand\@biblabel[1]{#1.} 
\makeatother

\usepackage{amssymb}
\usepackage[final]{pdfpages}
\usepackage{ucs}
\usepackage{subfigure}
\usepackage{color,graphicx}
\usepackage{amsmath}
\usepackage{amssymb}
\usepackage{amsxtra}
\usepackage{wrapfig}
\usepackage{setspace}
\usepackage{version}
\usepackage{fullpage}
\usepackage{framed}
\usepackage[color]{changebar}
\cbcolor{blue}
\usepackage{program}
\usepackage[numbers,round]{natbib}
\begin{document}

\noindent {\bf \LARGE The model muddle: in search of tumour growth laws}\\

\normalsize
\noindent Philip Gerlee\\
\small
\noindent Sahlgrenska Cancer Center, University of Gothenburg; Box 425, SE-41530 Gothenburg, Sweden \\
Mathematical Sciences, University of Gothenburg and Chalmers University of Technology, SE-41296 Gothenburg, Sweden
\normalsize
\section*{Abstract}
{\it In this article we shall trace the historical development of tumour growth laws, which in a quantitative fashion describe the increase in tumour mass/volume over time. These models are usually formulated in terms of differential equations that relate the growth rate of the tumour to its current state, and range from the simple one-parameter exponential growth model, to more advanced models that  contain a large number of parameters. Understanding the assumptions and consequences of such models is important, since they often underpin more complex models of tumour growth. The conclusion of this brief survey is that although much improvement has occurred over the last century, more effort and new models are required if we are to understand the intricacies of tumour growth.

}

\vspace{2cm}
\noindent The growth patterns exhibited by tumours have gathered the interest of scientists since the early days of cancer research, and despite almost a century of inquiry there are still uncertainties regarding the precise growth rate, or rather growth pattern, that solid tumours exhibit. This is in part due to the shift in scientific focus from the macroscopic increase in tumour size, to the (sub-)microscopic workings of signal transduction pathways, and the impact of specific genetic alterations. However, a lot can still be gained from an advance in our knowledge of tumour growth, not least in the clinic, where established growth curves may help to determine optimal time period between mammography screening \cite{Hart1998}, and may be used, as part of more complex models,  for calculating appropriate doses in radio- and chemotherapy \cite{Steel1989}. 

The increase in size of a neoplasm, caused by the up-regulation of cell division among malignant cells, is one of the most basic observations to be made when studying cancer \cite{Hanahan2000}. A natural question to ask is then how the size of the tumour increases over time as the disease progresses. What seems like a very simple question, has however turned out to be immensely difficult to answer, and the reasons why have only become evident in the last couple of decades, when many of the complex processes underlying tumour progression have been discovered. 

It was early on established that if cancer cells divide in a completely unconstrained fashion, and hence every cell in the tumour continuously passes through the cell cycle, and gives rise to two daughter cells at regular intervals, then the number of cancer cells and therefore the volume and mass of the tumour would increase exponentially with time. The geometric increase then implies that the time one has to wait for the tumour to double in size is constant over time. This picture of exponential growth has been shown to match the early stages of tumour growth, but in all known cases the doubling time eventually increases and continues to do so for the remainder of the disease. This can occur if either the average cell cycle time increases or if there is a loss of dividing cells due to quiescence or cell death. We shall discuss both these possibilities later on, but for now we simply note that other mechanisms need to be invoked.

\begin{figure}[!htb]
\begin{center}
\includegraphics[width=17cm]{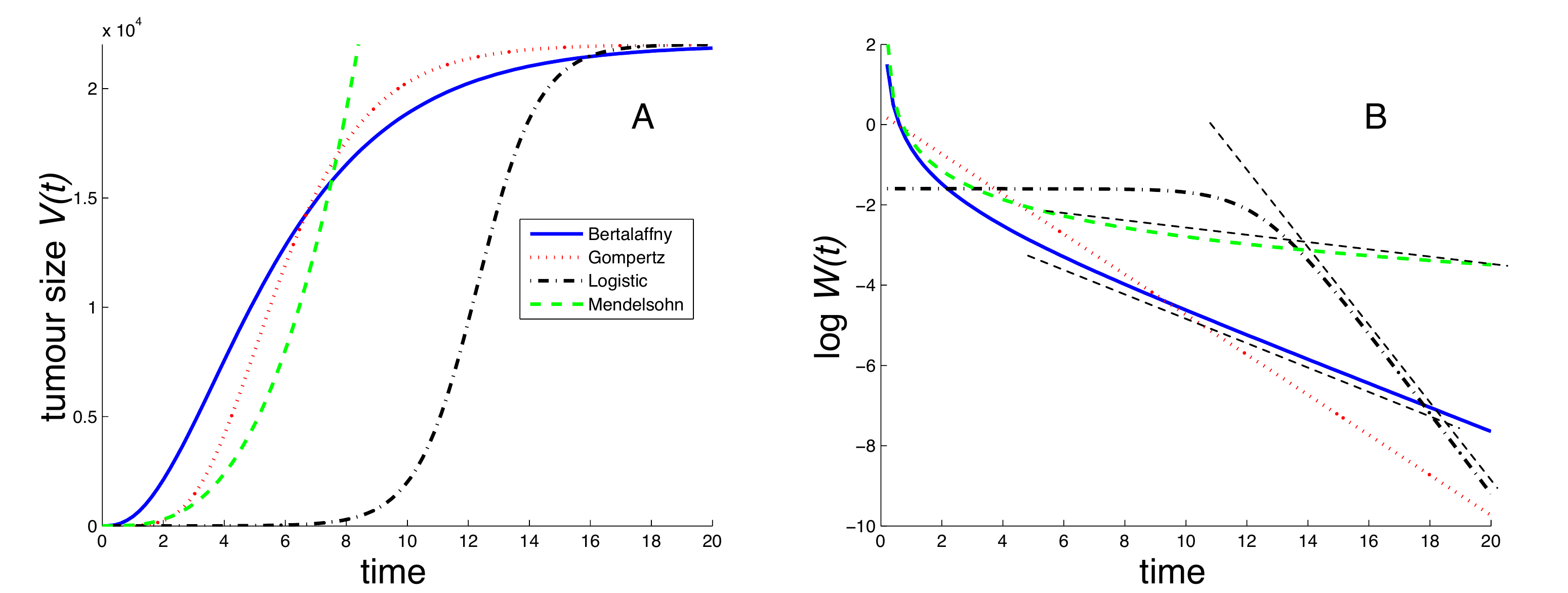}
\caption{\label{fig:fig}Tumour growth curves. Panel A shows the growth curves for four different models of tumour growth. The Gompertz (dotted), Bertalaffny (solid) and Logistic (dash dotted) tend to a asymptotic value, while the Mendelsohn model (dashed) gives rise to unconstrained growth. Panel B shows the quantity $\log \log V(t)/V(t-1)$ which is used for fitting the Gompertz model to experimental data. However, as the figure shows, the other models approximately exhibit a linear decrease in $\log W(t)$, which is usually considered to be a hallmark of Gompertzian growth.} 
\end{center}
\end{figure}

The paradigm of exponential growth is thus unable to explain growth dynamics of tumours in the longer term, and one has to look beyond the simple idea of unconstrained cell division to explain the observed data. This was indeed done in the first half of the 20th century through pioneering work by Mayenord (1932) \cite{Mayenord1932} and Schreck (1936) \cite{Schreck1936} among others, who considered models where the growth rate is retarded. However, in order to fully appreciate this advance it is necessary to resort to mathematical notation. Exponential growth is described by the differential equation
\begin{equation}\label{eq:exp}
\frac{dV(t)}{dt} = r V(t)
\end{equation}
which equates the rate of increase in volume $dV/dt$ to the current volume $V(t)$ times the growth rate $r$, assumed to be constant. This means that within an infinitesimal time interval $dt$ the increase in volume $dV$ is proportional to the current size of the tumour. The solution of this equation is $V(t)=V_0e^{rt}$, where $V_0$ is the volume of the tumour at time $t=0$ when measurements started, and $e$ is the base of the natural logarithm. Equation \eqref{eq:exp} can however be viewed as a special case of a more general equation
\begin{equation}\label{eq:expower}
\frac{dV(t)}{dt} = r V(t)^b
\end{equation}
which was introduced by Mendelsohn, where now the rate of increase is proportional to the volume raised to the power $b$, which can take on any value \cite{Mendelsohn1963}. Now if $b=1$ the solution is the above exponential growth curve, but for $b\neq1$ the solution is given by
\begin{equation}\label{eq:expowersol}
V(t)=\left( (1-b)(rt+C)\right)^{\frac{1}{1-b}}
\end{equation}
where $C$ is a constant related to the initial condition (see fig.\ \ref{fig:fig}A). By fitting growth curves from mouse mammary tumours to this equation, Dethlefsen et al. (1968)\ \cite{dethlefsen1968} were able to show that many tumours grow according to \eqref{eq:expower} with $b \approx 2/3$. This value of $b$ was in fact already suggested by Mayenord in 1932 \cite{Mayenord1932}, and can be derived from simple physical considerations: Assuming that the tumour is spherical in shape with volume $V$, its surface area scales as $V^{2/3}$. If we further assume that the growth of the tumour is limited by nutrients and/or oxygen which enter through the surface, then the growth rate of the tumour should be proportional to its surface area, i.e.\ $V^{2/3}$. In this case ($b=2/3$) the solution is given by $V(t) \sim t^3$, or in terms of the of the tumour radius $R(t)=\sqrt[3]{V(t)} \sim t$.

In other words the radius of the tumour grows linearly with time, and this behaviour can also be explained
by assuming that only a thin layer of cells at the surface of the tumour are in fact dividing. This suggestion was, at the time, highly disputed \cite{Patt1954, Baserga1960, Mendelsohn1962}, and results from animal models suggested that the whole tumour was mitotically active. 
In fact, for sub-exponential growth to occur, it is not sufficient for a constant fraction of the tumour cells to be quiescent, but an ever increasing fraction of cells must become mitotically inactive as the growth progresses. This seemed an unlikely scenario, given that tumours actually grow at a considerable rate, but we now know that this is the general mode of growth of solid tumours, at least prior to vascularisation: tumours contain a proliferating region of roughly constant width, and hence an ever diminishing fraction of active cells. 


Despite the improvement in describing tumour growth curves the above model could not account for the longer term reduction in growth rate often associated with later stages of the disease. This problem was addressed in a seminal paper by Laird (1963) \cite{Laird1963}, who suggested a fundamentally different model for explaining tumour growth, namely the Gompertz model, whose unexpected origin deserves a mention. It was put forward by Benjamin Gompertz in 1825 \cite{Gompertz1825} as a means to explain human mortality curves. Gompertz had made the observation that
``If the average exhaustions of a man's power to avoid death were such that at the end of equal infinitely small intervals of time, he lost equal portions of his remaining power to oppose destruction...'' then the number of living humans at age $x$ would be given by 
\begin{equation}\label{eq:gomp}
L(x) = k e^{-e^{a-bx}}
\end{equation}
where $k,a$ and $b$ are constants, of which $k$ and $b$ are necessarily positive. 

The main motivation for the model was actuarial, as a practical means of determining the value of life insurances, and only later was it proposed as a model for biological growth. This was done by the geneticist Sewall Wright in 1926 \cite{Wright1926}, who observed that ``the average growth power, as measured by the percentage rate of increase, tends to fall at a more or less uniform percentage rate''. Or in other words, the growth rate of an organism or organ tends to decrease at a constant rate.
The model was first applied by Davidson (1928) \cite{Davidson1928}, for describing the growth of cattle, who also gave the following derivation of the equation, which formalises what Wright had put in words. 

Assume that a growth process is governed by
\begin{equation}\label{eq:david1}
\frac{dV(t)}{dt} = r(t) V(t)
\end{equation}
where $V(t)$ is the mass or volume of an organism. This equation is similar to the equation of exponential growth \eqref{eq:exp}, but with the growth rate $r=r(t)$ now being time-dependent. We now assume that $r(t)$ decreases in proportion to its current value at a constant rate $\rho$, i.e.\ 
\begin{equation}\label{eq:david2}
\frac{dr(t)}{dt} = -\rho r(t).
\end{equation}
The solution of these two coupled equations now yields the Gompertz curve in its more familiar shape
\begin{equation}\label{eq:gomp2}
V(t) = V_0 e^{\frac{r_0}{\rho}(1-e^{-\rho t})}
\end{equation}
where $r_0$ is the growth rate at time $t=0$, and $V_0$ is the initial volume of the animal. This equation has successfully been fit to biological growth in a wide variety of contexts ranging the growth of internal organs \cite{Laird1965}, whole organisms \cite{Laird1965a} and entire populations \cite{Spickett1990}. In contrast to the exponential and Mendehlson-model, the growth curve generated by the Gompertz equation \eqref{eq:gomp2} is sigmoidal in shape, and reaches a constant value, or asymptote, as $t \rightarrow \infty$, with $V_\infty=V_0 e^{r_0/\rho}$ (see fig.\ \ref{fig:fig}A).

The previous success in describing biological growth is probably what motivated the application of the Gompertz equation to tumour growth, but Laird also gives some justification in terms of tumour biology. In arguing in favour of the Gompertz model Laird notes that if only a constant fraction of the tumour cells were cycling then this would still give rise to exponential growth, albeit with a smaller growth rate. The observed exponential decrease in overall growth rate coupled with experimental observations suggesting that almost all cancer cells in a tumour are passing through the cell cycle \cite{Laird1963} was instead readily explained by a model in which all cells cycled, but with an ever diminishing speed.

A suggested mechanism of action was growth retarding factors, whose effects increases during growth according to an exponential function, and one candidate was an accelerated immunological response \cite{Laird1963}. We now know that quite the opposite is true; large parts of solid tumours are quiescent, and the cells that are actually dividing,  do this at a rate comparative to the one at early stages of progression \cite{Knoefel1987}. This means that the decaying growth rate in \eqref{eq:david2} cannot be given any natural biological meaning, since it represents, not a single process, but the joint effect of many confounding factors, the total sum of which is negative.
In addition the doubling times required to match the early phases of tumor growth sometimes take on unrealistically small values ($<$ 10 hours) \cite{Steel1966}, suggesting fundamental problems with the modelling approach. Despite these drawbacks the Gompertz equation has proven to be a useful tool when describing tumour growth curves. It has been applied and varied in many different ways, for example to capture the effects of radiation \cite{Donoghue1997} and anti-angiogenic therapy \cite{Hahnfeldt1999}. The latter variation is interesting since it makes use of the notion of a variable maximal volume or `carrying capacity', imposed by some environmental limitation such as nutrients. In order to grasp this concept we need view the Gompertz equation in a different, but completely equivalent, form
\begin{equation}\label{eq:gomp3}
\frac{dV(t)}{dt} = \rho V(t) \log \frac{K(t)}{V(t)},
\end{equation}
where $K(t)$ is the carrying capacity of the tumour at time $t$, which in the previous time-independent equation would equal $V_\infty$. When $V(t)=K(t)$ we have $\log K(t)/V(t) = \log 1 = 0$, and the growth rate is equal to zero. Hahnfeldt et al.\ \cite{Hahnfeldt1999} coupled the carrying capacity to the angiogenic response by formulating a separate differential equation for $K(t)$, which takes into account the influence of the tumour mass on the dynamics of the vasculature. They could show that the model accurately describes the growth dynamics both in untreated tumours and those exposed to anti-angiogenic factors such as angiostatin.


The Gompertz model is however not the only model that can capture a decrease in tumour growth rate over time and an asymptotic mass, and at least two other models have been put forward as plausible candidates. The logistic (or Pearl-Verhulst) equation given by
\begin{equation}\label{eq:log}
\frac{dV(t)}{dt} = rV(t)\left(1-\frac{V(t)}{K}\right).
\end{equation}
was first formulated by Pierre Fran\c{c}ois Verhulst in 1838 \cite{Verhulst1838} as a means of describing the dynamics of a population with an intrinsic growth rate $r$, whose total size is limited by a carrying capacity $K$ (see fig.\ \ref{fig:fig}A). It has since then become a mainstay of biomathematics and has successfully been applied to a large number of biological phenomena, ranging from bacterial populations to algae and mammals \cite{Murray1989}. Whereas the Gompertz equation assumes an exponentially decreasing growth rate, the logistic equation instead assumes that the growth rate falls off linearly with the size, until it becomes equal to zero when it reaches the carrying capacity $V(t)=K$. In terms of matching actual growth curves, the logistic and Gompertz equation are quite similar \cite{Winsor1932}, with the main difference being that the Gompertz curve is asymmetric, with the point of inflection (the time point where the growth rate is maximal) occurring after 37 \% of the final size has been reached, while for the logistic this occurs after half of the growth has occurred. The logistic equation can be derived by considering a spatially extended population where reproduction is constrained by available space. Although the competition for space plays an important role in tumour growth, 
this derivation ignores many other important factors, such as limited nutrients, that also influence the process. This leaves the model hanging in a phenomenological void, also inhabited by the Gompertz model, with the possible advantage that it has a mechanistic derivation.

A second candidate, which has received considerably less attention is the Bertalanffy equation  
\begin{equation}\label{eq:vbf}
\frac{dV(t)}{dt} = aV(t)^{2/3}-bV(t)
\end{equation}
whose instigator was the founder of general systems theory Ludwig von Bertalanffy. It was put forward as a model for organism growth \cite{Bertalanffy1949}, and its derivation is similar to the above model by Mendelsohn \cite{Mendelsohn1963}, equation \eqref{eq:expower}, i.e. growth occurs proportional to surface area, with the additional assumption that the loss of tumour mass due to cell death occurs in proportion to the volume of the tumour with a constant $b$, related to the commonly employed cell loss factor. The solution of \eqref{eq:vbf} is also sigmoidal in shape, and tends to a fixed volume as time increases, where the growth and loss term balance each other out (see fig.\ \ref{fig:fig}A). The striking thing about the Bertalanffy equation is that it both matches experimental tumour growth curves well, and in addition has a derivation with biologically meaningful parameters. In a review of different tumour growth models it was in fact shown that the Bertalaffny model gave a better fit than both the Gompertz and Logistic model in seven out of ten cases \cite{Vaidya1982}. %

Despite this fact the Gompertz model has remained the most applied models when it comes to describing tumour growth curves (see for example Norton (1988) \cite{Norton1988}). How can this be the case? We believe the reasons are two-fold, one theoretical and one practical. 

The former reason is related to the nature and purpose of modeling. Do we formulate and apply a model in an effort to understand a system, or in order to predict its future behaviour? The answer is rarely clear cut, but if one leans towards prediction then a model which is disconnected from reality in terms of mechanisms and dynamics is acceptable, as long as it does the job of predicting. If one, on the other hand, has a yearning for understanding the system at hand, then the model has to be derived from, and based on, real mechanisms and entities within the system. Since the purpose of growth curves is often to predict the future size of the tumour, it is not surprising that the phenomenological Gompertz model has dominated. 

The second reason is of a more practical nature, and connected to the method with which one fits experimental data to the Gompertz model. In order to find the parameters of the model one forms the quantity $W(t)=\log V(t) - \log V(t-1)$ which decays exponentially with time according to $W(t) = Ae^{-\rho t}$, where $A=r_0/\rho(e^\rho-1)$. Thus in order to find the parameters of the model one plots the quantity $\log W(t) = \log A -  \rho t$ as a function of time, and by applying some regression method, such as least squares, finds $\rho$ as the (negative) slope and $\log A$ as the intercept. In this process we essentially take the logarithm of the tumour volume $V(t)$ twice, suppressing any deviations from the Gompertz model, and thus making it quite easy to fit the model to data that follows a completely different growth law. This fact is illustrated in fig. \ref{fig:fig}B, which shows four different growth curves (Gompertz, Logistic, Mendelsohn and Bertalaffny) and the quantities $\log W(t)$ plotted over time. It is evident that even though the growth curves $V(t)$ look quite different, the plots of $\log W(t)$, at least for longer times, are very close to being straight lines, which is consistent with the Gompertz model. In the case of real tumour data with its inherent experimental error, and in some cases late detection, this distinction is even harder to make, and in one sense the Gompertz model is poised to win the race. 

In my opinion these are the two main reasons why the Gompertz model has held a dominating position for nearly half a century. However, the ease with which a model can be fitted to data, should not determine its use over other, more accurate models, and also the latter methodological consideration we find worth scrutinising. A model which is grounded in the actual biology (which the Gompertz model is not) can not only provide predictions, but also give insight into the underlying dynamical process of tumour growth. By comparing the parameter values obtained from different patients or cell lines, one might be able to draw conclusions about the basis of growth and its dynamics. Among the candidates we have reviewed here it seems to me that the Bertallanffy equation of tumour growth is the most suitable candidate. It produces growth curves which are nearly indistinguishable from the well-known Gompertz model, but has the advantage of being biologically motivated. 

However, more research is needed in this field, which for a long time has remained dormant. A notable exception to this is the model by Herman et al. (2011) \cite{Herman2011}, which takes into account the tumour vascular network and its interface with the cardiovascular system of the host. This comprehensive model relates tumor vascularisation and growth to metabolic rate, and gives predictions for tumour properties. These include growth rates, metabolic rates, degree of necrosis, blood flow rates and vessel sizes, all of which are clinically relevant variables, and shows that models such as this one are helpful in our future inquiry into the dynamics of tumour growth.

%
%

\end{document}